\documentclass[pra,twocolumn, prl,groupedaddress]{revtex4-1}

\usepackage{graphicx}
\usepackage{dcolumn}
\usepackage{bm}
\usepackage{epsfig}
\usepackage{amsmath}
\usepackage{amssymb}
\usepackage{color}
\usepackage{bbm}

\usepackage[colorlinks=true,citecolor=blue,linkcolor=blue,urlcolor=blue]{hyperref}

\begin{document}

\title{Intracity quantum communication via thermal microwave networks}

\author{Ze-Liang~Xiang$^{1}$, Mengzhen~Zhang$^{2,3}$,  Liang Jiang$^{2,3}$, and Peter~Rabl$^1$}
\affiliation{$^1$Vienna Center for Quantum Science and Technology, Atominstitut, TU Wien, Stadionallee 2, 1020 Vienna, Austria}
\affiliation{$^2$Departments of Physics and Applied Physics, Yale University, New Haven, Connecticut 06520, USA}
\affiliation{$^3$Yale Quantum Institute, Yale University, New Haven, Connecticut 06520, USA}

\date{\today}

\begin{abstract} 
Communication over proven-secure quantum channels is potentially one of the most wide-ranging applications of currently developed quantum technologies. It is generally envisioned that in future quantum networks, separated nodes containing stationary solid-state or atomic qubits are connected via the exchange of optical photons over large distances. In this work we explore an intriguing alternative for quantum communication via all-microwave networks. To make this possible, we describe a general protocol for sending quantum states through thermal channels, even when the number of thermal photons in the channel is much larger than 1. The protocol can be implemented with state-of-the-art superconducting circuits and enables the transfer of quantum states over distances of about 100 m via microwave transmission lines cooled to only $T=4$ K. This opens up new possibilities for quantum communication within and across buildings and, consequently, for the implementation of intra-city quantum networks based on microwave technology only. 
\end{abstract}

\maketitle

%
%


\maketitle

\section{Introduction}

Superconducting circuits~\cite{Schoelkopf2008,Clarke2008,You2011} are considered one of the most promising platforms for implementing quantum information processing schemes, where all the key elements, like high-fidelity single- and two-qubit gates~\cite{Barends14}, efficient readout~\cite{Vijay11,Sun2014}, and error correction capabilities~\cite{Kelly15,Ofek16,Takita16} have already been experimentally demonstrated. However, superconducting qubits are usually operated at transition frequencies of a few GHz, which requires cooling of the circuits to a few tens of mK in order to avoid  detrimental thermal excitations. In practice, this restricts the generation of entanglement and the coherent exchange of quantum information to qubits and photons located within the same dilution refrigerator~\cite{Eichler2012,Menzel2012,Roch2014,Flurin2015}. To overcome this limitation, various hybrid system approaches for coherently interfacing microwave and optical photons are currently explored~\cite{Xiang2013,Kurizki2015}. Establishing such interfaces with, e.g., optomechanical systems~\cite{WangPRL12,Tian12,Barzanjeh2012,Bochmann2013,Andrews2014,Bagci2014,Rueda2016,Tian2015}, atoms~\cite{Verdu2009,Hafezi12}, or spin ensembles~\cite{Kubo2011,Zhu11,Probst2013,Tabuchi15,Putz16,Hisatomi2016,OBrien2014} would provide access to optical long-distance quantum communication~\cite{Cirac1997,Kimble2008,Ritter2012} but naturally comes at the price of introducing additional experimental overheads and sources of decoherence.

\begin{figure}[t]
  \centering
    \includegraphics[width=0.48\textwidth]{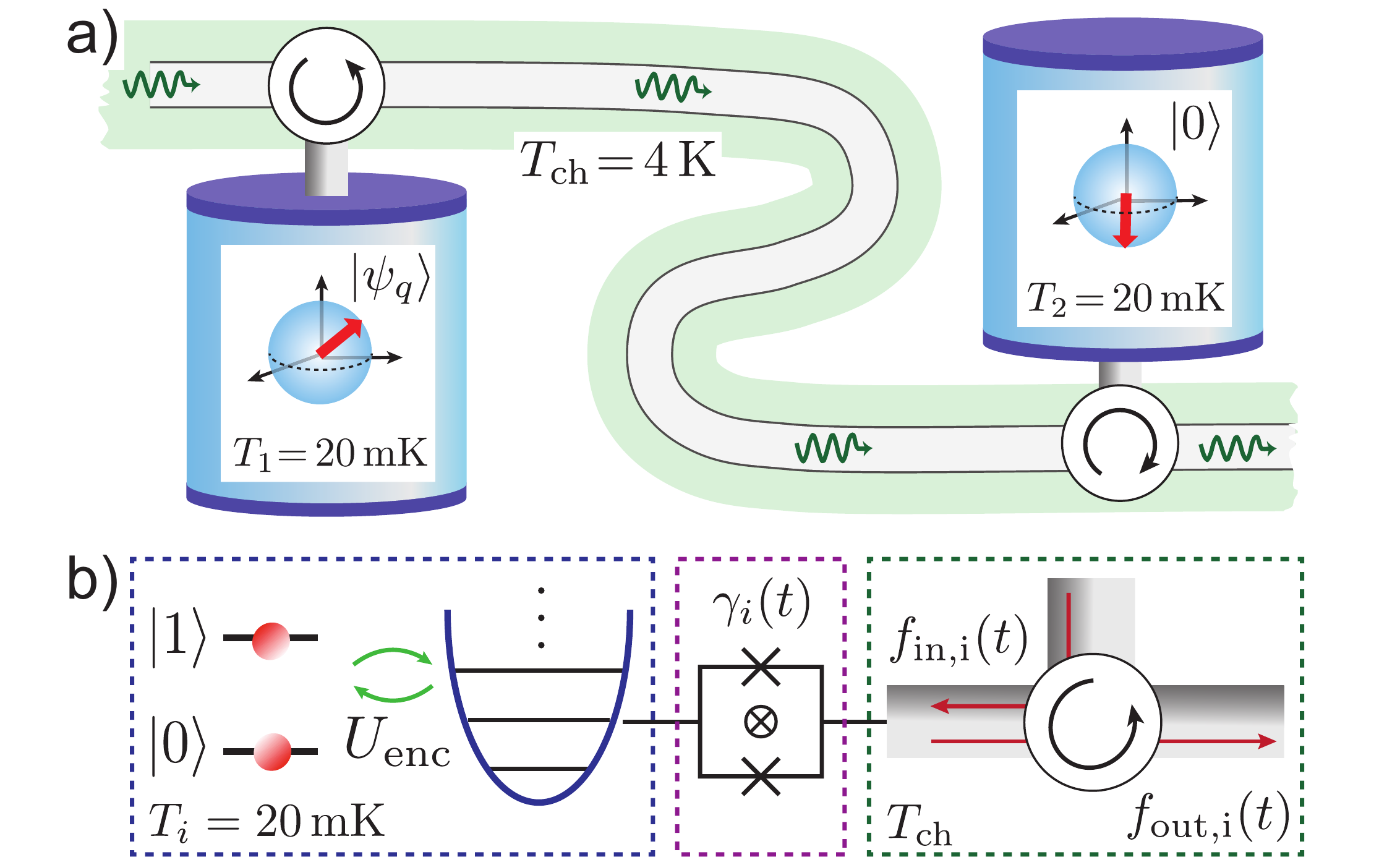}
      \caption{(a) Sketch of a thermal quantum network, where two nodes (for example, two superconducting qubits located inside separated dilution refrigerators) are connected via a unidirectional quantum communication channel at finite temperature $T_{\rm ch}$. (b) For the implementation of a noise-resilient transfer protocol, the qubit state is first mapped onto an intermediary oscillator. The oscillator is then coupled to the incoming and outgoing fields of the channel, $f_{\rm in,i}(t)$ and $ f_{\rm out,i}(t)$, via a tunable decay rate $\gamma_i(t)$, which can be realized, for example, by a flux-tunable quantum interference device~\cite{Korotkov2011,Yin13,Wulschner16}.}
      \label{Fig1Setup}
\end{figure}

In this work we discuss an interesting alternative approach for superconducting quantum communication, which avoids  technical challenges associated with optical interfaces and pursues the direct transmission of quantum states via thermal microwave channels. To make this idea feasible in practice, we describe a conceptually simple, but at the same time very powerful strategy, which enables a \emph{perfect} transfer of quantum states through a thermal channel, even if the average number of thermal photons in the channel is much larger than 1. This becomes possible by introducing at each node an intermediary oscillator as a controllable port between the finite-dimensional qubits and the bosonic channel. By making use of the large Hilbert space provided by the oscillator, one is further able to implement error-correction protocols~\cite{Bennett1996,Knill1997}, which correct for both photon loss as well as photon absorption errors~\cite{Michael2016}, as relevant in a finite temperature environment. This makes the protocol robust against pulse imperfections and losses encountered in a realistic scenario. 

While already on a chip-scale various phononic~\cite{SafaviNaeiniNJP2011,Habraken2012,GustafssonScience2014,SchuetzPRX2015}, microwave or hybrid quantum networks could benefit from the following protocol, a key application lies in the coherent coupling of superconducting qubits located in different dilution refrigerators, via superconducting transmission lines held at a much more convenient temperature of $T\approx 4$ K. Our analysis shows that by using already-existing superconducting circuit technology combined with realistically achievable loss rates in  microwave transmission lines, a deterministic exchange of quantum information over tens and even hundreds of meters is possible. At this threshold, communication across buildings and consequently, the establishment of fully connected microwave quantum networks within densely populated areas are within technological reach.

\section{Quantum state transfer}

Figure~\ref{Fig1Setup} shows a minimal instance of a quantum network, where two nodes $i=1,2$ are connected via a unidirectional quantum communication channel. Each node is represented by a single well-isolated quantum system, for example,  a superconducting two-level system, with a ground state $|0\rangle$ and a first excited state $|1\rangle\equiv L^\dag |0\rangle$ with creation operator $L^\dagger$, which are separated in frequency by $\omega_0$. 
At time $t=t_0$, the qubit in node 1 is prepared in a pure quantum state $|\psi_q\rangle = \alpha|0\rangle+\beta|1\rangle$, and the second qubit is in its ground state. 
A fundamental task in a quantum network is to transfer the state $|\psi_q\rangle$ from node 1 to node 2 via a tunable  coupling to the quantum channel. Starting from the density operator of the whole network, $\rho(t_0)= | \psi_q\rangle_1\langle \psi_q|\otimes  \rho_{\rm ch}(t_0) \otimes |0\rangle_2\langle 0| $, where $\rho_{\rm ch}(t_0)$ is the initial state of the quantum channel, this corresponds to a mapping 
\begin{equation}
\begin{split}
\rho(t_0)\rightarrow  \rho(t_f)=\rho_{1\&{\rm ch}}(t_f)  \otimes | \psi_q\rangle_2\langle \psi_q|,
 \end{split}
\end{equation}
where $\rho_{1\&{\rm ch}}(t_f)$ denotes the combined states of node 1 and the channel at the final time $t_f$. Under realistic conditions this transfer will never be perfect and we use the average fidelity~\cite{Nielsen02}, 
\begin{equation}\label{eq:Fidelity}
\bar{\mathcal{F}}= \int d\psi_q \,  {\rm Tr}\{| \psi_q\rangle_2\langle \psi_q |  \rho(t_f) \} \leq 1,
\end{equation}
as a measure for the quality of the state transfer protocol. Above a minimum of $\bar{\mathcal{F}}>2/3$ the transfer is genuine quantum, i.e., it cannot be reproduced by measurements and classical communication~\cite{Massar1995}. 

\begin{figure}[t]
  \centering
    \includegraphics[width=0.48\textwidth]{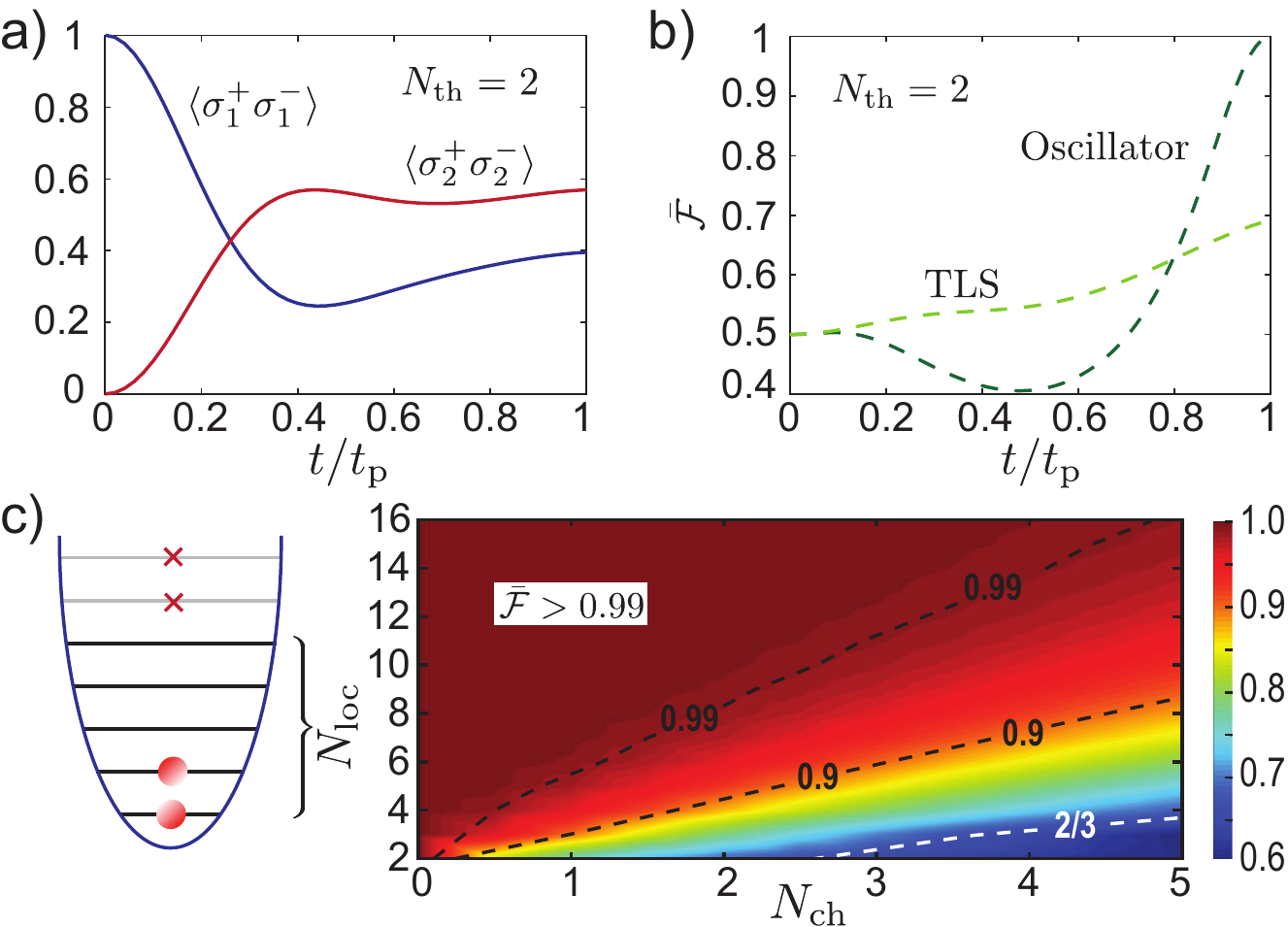}
      \caption{(a) Occupation number of qubits 1 and 2 ($L_i\equiv \sigma_i^-$) during the evolution generated by $H_{\rm int}$ in Eq.~\eqref{eq:HintSimple} 
      and for a thermal occupation $N_{\rm ch}=2$ of the channel mode. (b) Comparison of the average fidelity $\bar{\mathcal{F}}$ for quantum states encoded locally in a two-level system (TLS) and a harmonic oscillator. (c) Average transfer fidelity for the case, where the dynamics of the local oscillators is restricted to the $N_{\rm loc}$ lowest number states.}
      \label{Fig2}
\end{figure}

To illustrate the basic idea behind a noise-resilient transfer protocol, it is instructive to first consider a simple toy model, where the channel is represented by a single bosonic mode with annihilation operator $c$ and frequency $\omega_{\rm ch}=\omega_0$. Given that this mode is initially in the vacuum state, i.e., $|\Psi(t_0)\rangle=|\psi_q\rangle_1|0\rangle_{\rm ch}|0\rangle_2$, a state transfer can be achieved by applying the coupling 
\begin{equation}\label{eq:HintSimple}
H_{\rm int} = \hbar g\left[ (L_1+L_2)c^\dag + c (L_1^\dag +L_2^\dag)\right],
\end{equation}
for a time $t_{\rm p}=t_f-t_0=\pi/(\sqrt{2}g)$. Then, the unitary evolution $U(t)=\exp(-iH_{\rm int} t/\hbar)$ implements the desired operation
\begin{equation}
U(t=t_{\rm p})|\Psi(t_0)\rangle= |0\rangle_1\otimes |0\rangle_{\rm ch} \otimes |\psi_q\rangle_2,
\end{equation}
with perfect fidelity. However, the same process fails if the channel is initially in a thermal state with a temperature $T_{\rm ch}> \hbar \omega_0/k_B$ and, therefore, a random distribution of number states $|n\rangle_{\rm ch}$. In this case the second qubit can absorb either a photon from the first node or simply a thermal photon already present in the channel.  As shown in Figs.~\ref{Fig2}(a) and \ref{Fig2}(b), the resulting transfer fidelity drops significantly for an equilibrium occupation number $N_{\rm ch}=(e^{\hbar\omega_0/k_BT_{\rm ch}}-1)^{-1} \gtrsim 1$. 

Surprisingly, this seemingly unavoidable thermal degrading of quantum channels is not fundamental and is mainly a problem of interfacing the two-level qubit directly with the infinite-dimensional bosonic channel. If instead the initial and final quantum states are encoded in the lowest levels of a harmonic oscillator, where $L_{i}\equiv a_i$ are bosonic annihilation operators, i.e., $[a_i,a_j^\dag]=\delta_{ij}$, the same evolution from above implements the swap operation
\begin{equation}\label{eq:Uthermal}
U(t_{\rm p})|\psi_q\rangle_1\otimes |n\rangle_{\rm ch}\otimes  |0\rangle_2 = |0\rangle_1 \otimes |n\rangle_{\rm ch} \otimes |\psi_q\rangle_2,
\end{equation}
independently of the initial number state $|n\rangle_{\rm ch}$. In other words, the strict linearity of this simple toy model ensures a perfect state transfer fidelity, independently of the channel temperature [cf. Fig.~\ref{Fig2}(b)]. To further illustrate this point, we also consider in Fig.~\ref{Fig2}(c) the case, where the dynamics of each node oscillator is restricted to states $|n\rangle_i$ with $n< N_{\rm loc}$. We see that, for a given $N_{\rm ch}$, the linearity of the local quantum systems must be preserved up to $N_{\rm loc}\approx 3 N_{\rm ch}$ to achieve fidelities of $\bar F\simeq 0.99$. More generally, this example shows that, in contrast to the $T_{\rm ch}=0$ case, a quantum state  transfer at finite temperatures will strongly depend on the Hilbert space dimension of the local system that is used to encode the quantum state.

\section{Long-distance quantum communication}

The transfer of quantum states through thermal communication channels has been discussed previously, for example,  in the context of harmonic oscillator~\cite{Yao2013} and spin chains~\cite{Yao2013,Yao2011} of finite length or in multimode optomechanical systems~\cite{WangPRL12,Tian12}. These schemes either rely on fine-tuned couplings or the existence of a single isolated channel mode~\cite{Yao2013}, which can be resonantly addressed to achieve a situation similar to the one discussed above. The key observation is that the basic mechanism that underlies the coherent cancellation of noise in finite-size systems can be generalized to a continuum scenario, as is relevant for long-distance quantum communication. This suggests the general three-step protocol involving a two-level qubit and an additional intermediary oscillator at each node [cf. Fig. 1 b)]. In a first step, the state of the qubit is mapped onto a superposition of two oscillator states, i.e., $|\phi_0\rangle=|n=0\rangle$ and $|\phi_1\rangle=|n=1\rangle$, by a unitary encoding operation, $U_{\rm enc}|\psi_q \rangle \otimes|0\rangle=|0\rangle \otimes (\alpha |\phi_0\rangle+\beta|\phi_1\rangle)$. In a second step, the qubits and the oscillators are decoupled, and a state transfer between the oscillators via the thermal channel is implemented. Finally, the encoding operation is reversed in the second node. 

For the analysis of the state transfer between the intermediary oscillators, we consider a unidirectional quantum channel represented by a continuum of right-propagating bosonic modes. Each node is coupled to the channel by a controllable interaction of the form  
\begin{equation}\label{eq:Hint}
H_{\rm int} (t)= i \hbar \sum_i\sqrt{\gamma_i(t)}\left(a_if_{\rm ch}^\dag(z_i,t)-a_i^\dag f_{\rm ch}(z_i,t)\right),
\end{equation}
where $z_i$ denotes the position of node $i$ along the channel. In Eq.~\eqref{eq:Hint}, $f_{\rm ch}(z_i,t)=1/\sqrt{2\pi}\int_{\omega_0-\Delta}^{\omega_0+\Delta} d\omega f_\omega e^{-i\omega(t-z_i/v_{\rm g})}$ is the operator for the right-propagating field in the channel, where $v_{\rm g}$ is the group velocity and $f_\omega$ is the bosonic operator for the plane wave mode normalized to $[f_\omega,f_{\omega'}^\dag]=\delta(\omega-\omega')$. In the limit where the bandwidth $\Delta$ is sufficiently large, this coupling gives rise to a Markovian decay of the oscillators with tunable rates $\gamma_i(t)$. In a frame rotating with $\omega_0$, the dynamics of the whole network is then well described by a set of quantum Langevin equations for the Heisenberg operators $a_i(t)$~\cite{Gardiner1993,Carmichael1993,Gardiner2000},
\begin{equation}\label{eq:QLE}
\dot a_i(t) = -\frac{\gamma_i(t)}{2} a_i(t)  - \sqrt{ \gamma_i(t)} f_{\rm in,i}(t),
\end{equation}
together with the input-output relations $f_{\rm out,i}(t) =  f_{\rm in,i}(t)+ \sqrt{\gamma_i(t)} a_i(t)$. Here, $f_{\rm in,i}(t)=f_{\rm ch}(z_i+0^-,t)$ and $f_{\rm out,i}(t)=f_{\rm ch}(z_i+0^+,t)$ represent the field of the channel right before and after interacting with node $i$. For the first node, $f_{\rm in, 1}(t)\equiv f_{\rm in}(t)$ is the unperturbed field satisfying $\langle f_{\rm in}^\dag(t)f_{\rm in}(t')\rangle=\delta(t-t')N_{\rm ch}$ and $\langle f_{\rm in}(t)f_{\rm in
}^\dag(t')\rangle=\delta(t-t')(N_{\rm ch}+1)$. The incoming field at the second node is $f_{\rm in,2}(t)=f_{\rm out,1}(t-\tau)$, or
\begin{equation}\label{eq:fin2}
f_{\rm in,2}(t)= f_{\rm in}(t-\tau)+ \sqrt{\gamma_1(t-\tau)} a_1(t-\tau).
\end{equation}
Without loss of generality, we can set the retardation time $\tau=(z_2-z_1)/v_{\rm g}$ to zero, which amounts to a redefinition of all operators and pulses in the first node, i.e., $a_1(t-\tau)\rightarrow a_1(t)$, $\gamma_1(t-\tau)\rightarrow \gamma_1(t)$, etc.

Combining Eqs.~\eqref{eq:QLE} and \eqref{eq:fin2}, we obtain a coupled set of linear differential equations for the operators $a_i(t)$ with a general solution~\cite{Jahne2007} (see Appendix~\ref{app:Solution})
\begin{eqnarray} \label{eq:c1}
a_1(t)&=& A_1(t,t_0) a_1(t_0) + F_1(t,t_0),\\
a_2(t)&=& A_2(t,t_0) a_2(t_0) +T(t,t_0) a_1(t_0) + F_2(t,t_0). \label{eq:c2}
\end{eqnarray}
Here, $A_i(t,t_0)=e^{-\Gamma_i(t,t_0)/2}$ accounts for the decay of the initial amplitudes with an integrated loss $\Gamma_i=\int_{t_0}^t ds\, \gamma_i(s)$,
\begin{equation}\label{eq:T}
T(t,t_0)=-\int_{t_0}^t ds  \, A_2(t,s)
\sqrt{\gamma_1(s)\gamma_2(s)}   A_1(s,t_0)
\end{equation}
is the transfer amplitude and the $F_i(t,t_0)=\int_{t_0}^t ds \,  D_i(t,s) f_{\rm in}(s)$ represent the accumulated noise in each node. For the first node, $D_1(t,s)=-\sqrt{\gamma_1(s)}A_1(t,s)$, while for the second node, which we are primarily interested in, the filter function is  
\begin{equation}\label{eq:D2}
D_2(t,s) = -T(t,s)\sqrt{\gamma_1(s)} - A_2(t,s)\sqrt{\gamma_2(s)}.
\end{equation}
It reflects the interference between noise that is driving node 2 directly and noise that was first absorbed and then reemitted by node 1. Equations.~\eqref{eq:c1} and~\eqref{eq:c2} are the general results for a state transfer process in a unidirectional resonator network. For zero temperature, where such processes have already been analyzed in more detail~\cite{Jahne2007,Korotkov2011,Sete2015}, the noise terms $F_{1,2}$ do not contribute. In this case, these results are also identical to the corresponding equations for the state transfer amplitudes of coupled two-level systems~\cite{Cirac1997,StannigelPRA2011,Habraken2012}. At finite temperature, this correspondence is lost.

From Eq.~\eqref{eq:c2}, it follows that a perfect state transfer is achieved if, at some final time, $|T(t_f,t_0)|\rightarrow 1$, while  $A_{2}(t_f,t_0)\rightarrow0$, $D_2(t_f,t_0)\rightarrow 0$. To see that there are indeed pulses $\gamma_i(t)$ that can achieve these conditions, one can first look at  the known case $N_{\rm ch}=0$, where a perfect state transfer also implies that no photon leaves the system to the right, i.e., $f_{\rm out,2}(t)|\Psi(t_0)\rangle=0, \forall t$~\cite{Cirac1997}.  From this observation and using $f_{\rm in}(t)|\Psi(t_0)\rangle=a_2(t_0)|\Psi(t_0)\rangle=0$, the dark state condition
\begin{equation}\label{eq:DarkState}
A_1(t,t_0)\sqrt{\gamma_1(t)}  + T(t,t_0)\sqrt{\gamma_2(t)} =0,
\end{equation} 
can be derived. It relates $\gamma_1(t)$ and $\gamma_2(t)$ at each point in time and can be used to explicitly construct optimal transfer pulses~\cite{Cirac1997,StannigelPRA2011}.  In the presence of thermal photons such a physical argument no longer applies. However, one can show instead that condition~\eqref{eq:DarkState} also implies that the quantity $\mathcal{N}(t)=|A_1(t,t_0)|^2 + |T(t,t_0)|^2$ is conserved~\cite{Habraken2012}, i.e., $\partial_t \mathcal{N}(t)=0, \,\forall t$.  In view of $A_1(t_0,t_0)=1$ and $A_1(t_f,t_0)\rightarrow 0$, this ensures $|T(t_f,t_0)|\rightarrow1$ also for a noisy channel. Finally, we can use the commutation relation $[a_2(t),a_2^\dag (t)]=1$ to obtain the relation 
\begin{equation}
1 = |A_2(t,t_0)|^2 + |T(t,t_0)|^2 + \int_{t_0}^t ds  |D_2(t,s)|^2,
\label{Eq13}
\end{equation}
and prove that for $|T(t_f,t_0)|^2=1$, $|A_2(t_f,t_0)|^2=0$ and $D_2(t_f,s)=0$, $\forall s\in[t_0,t_f]$ are automatically fulfilled. Therefore, we have shown that although the dynamics of a thermal network is much more involved during the process, a perfect state transfer can still be implemented using the same control pulse as is applicable for a vacuum channel. This result is independent of the amount of thermal or other types of noise, but it requires a strict linearity of the intermediary oscillators, up to a number of excitations $N_{\rm loc}$ estimated above.

\begin{figure}[t]
  \centering
    \includegraphics[width=0.48\textwidth]{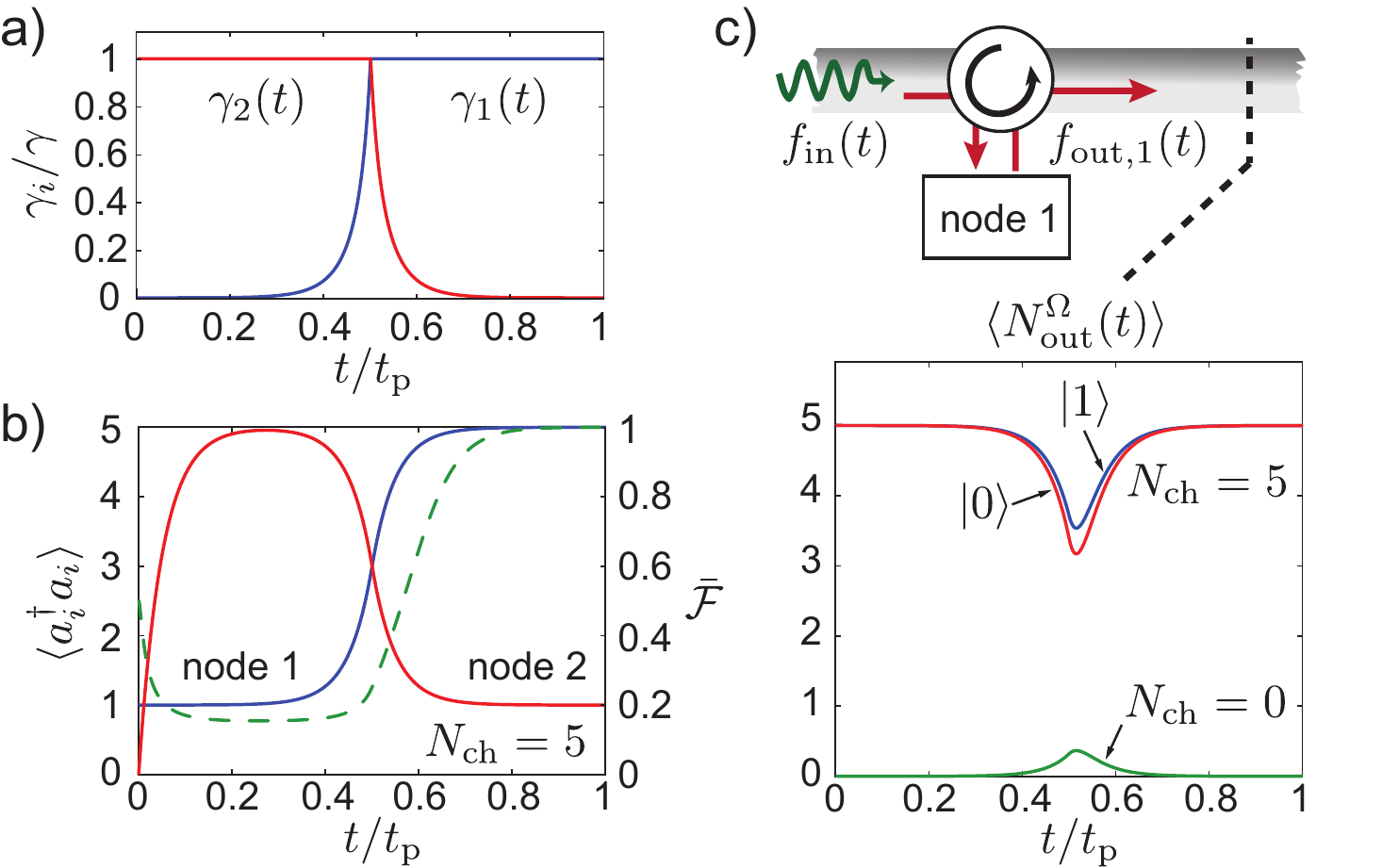}
      \caption{(a) Example control pulses $\gamma_{i}(t)$ for implementing a perfect state transfer over long distances via a thermal quantum communication channel. Here, $\gamma_1(t)=\gamma e^{\gamma (t-t_{\rm p})}/(2-e^{\gamma (t-t_{\rm p})})$ for $t<t_{\rm p}/2$ and $\gamma_1(t)=\gamma$ for $t\geqslant t_{\rm p}/2$~\cite{StannigelPRA2011}. The recapture pulse $\gamma_2(t)$ is chosen to be symmetric with respect to $t=t_{\rm p}/2$, and for all plots, a total pulse duration of  $t_{\rm p}=20\gamma^{-1}$ has been assumed. (b) The resulting evolution of the populations of each intermediary oscillator for an initial state $|\psi_q\rangle=|1\rangle$. The green dashed line shows the average transfer fidelity for the same pulse sequence. (c) Expectation value of the photon number in the channel measured in between node 1 and node 2 with  a detector of bandwidth $\Omega/\gamma=4$. The individual curves show the results for $N_{\rm ch}=0$ and $|\psi_q\rangle=|1\rangle$ (green line), $N_{\rm ch}=5$ and $|\psi_q\rangle=|1\rangle$ (blue line) and  $N_{\rm ch}=5$, and $|\psi_q\rangle=|0\rangle$ (red line).}
      \label{Fig3}
\end{figure}

Figure~\ref{Fig3} summarizes the results of the transfer protocol for a specific example of a time-symmetric pulse, which implements a quantum state transfer with a residual error  $\varepsilon_{\rm p}\sim e^{-\gamma t_{\rm p}/2}$, where $\gamma$ is the maximal decay rate. The evolution of the populations plotted in Fig.~\ref{Fig3}(b) clearly demonstrates that while the second node quickly thermalizes at the beginning of the process, all the noise is completely rejected at the end of the pulse. In Fig.~\ref{Fig3}(c) we also show the outcome of a (hypothetical) measurement of the photon number in the channel between node 1 and node 2. Here, we have defined $\langle N_{\rm out}^\Omega(t)\rangle = \langle [f_{\rm out}^\Omega(t)]^\dag f_{\rm out}^\Omega(t)\rangle$, where $f_{\rm out}^\Omega(t)=\sqrt{\Omega}\int_{-\infty}^t ds\,  e^{-\frac{\Omega}{2}(t-s)}f_{\rm out,1}(s)$, and $\Omega$ is the bandwidth of the detector. Under these conditions where one is interested in observing the evolution of the transmitted state, i.e., $\Omega\gg \gamma$, we obtain  
\begin{equation}
\begin{split}
\langle N_{\rm out}^\Omega(t)\rangle\simeq  &\frac{4\gamma_1(t)}{\Omega} |A_1(t,t_0)|^2 \langle a^\dag_1(t_0)a_1(t_0)\rangle
\\&+ N_{\rm ch}\left[ 1- \frac{4\gamma_1(t)}{\Omega} |A_1(t,t_0)|^2\right].
\end{split}
\end{equation}
This result demonstrates another interesting point, namely, that at every instance in time, the actual photon number in the channel is much larger than the amplitude of the transmitted signal photon. Compared to the case of a vacuum channel, it is thus not possible to detect the transmitted photon, without knowing the precise shape of the recovery pulse.

\section{Imperfections and error correction}
In our analysis so far, we have considered an arbitrary initial thermal occupation of the channel but assumed otherwise ideal conditions. To assess the actual practicability of the protocol, we must evaluate its robustness with respect to various sources of imperfections.  By restricting the following discussion to imperfections related to the transfer itself,  we observe that, first of all, imprecisions in the control pulses lead to an incomplete transfer, $|T(t_f,t_0)|<1$. As a consequence, the rejection of noise will also be incomplete. Second, any losses in the channel will not only reduce the signal but are also necessarily accompanied by additional thermal noise, which is uncorrelated with $f_{\rm in}(t)$ and therefore cannot be coherently canceled. 

\begin{figure}[t] 
  \centering
    \includegraphics[width=0.48\textwidth]{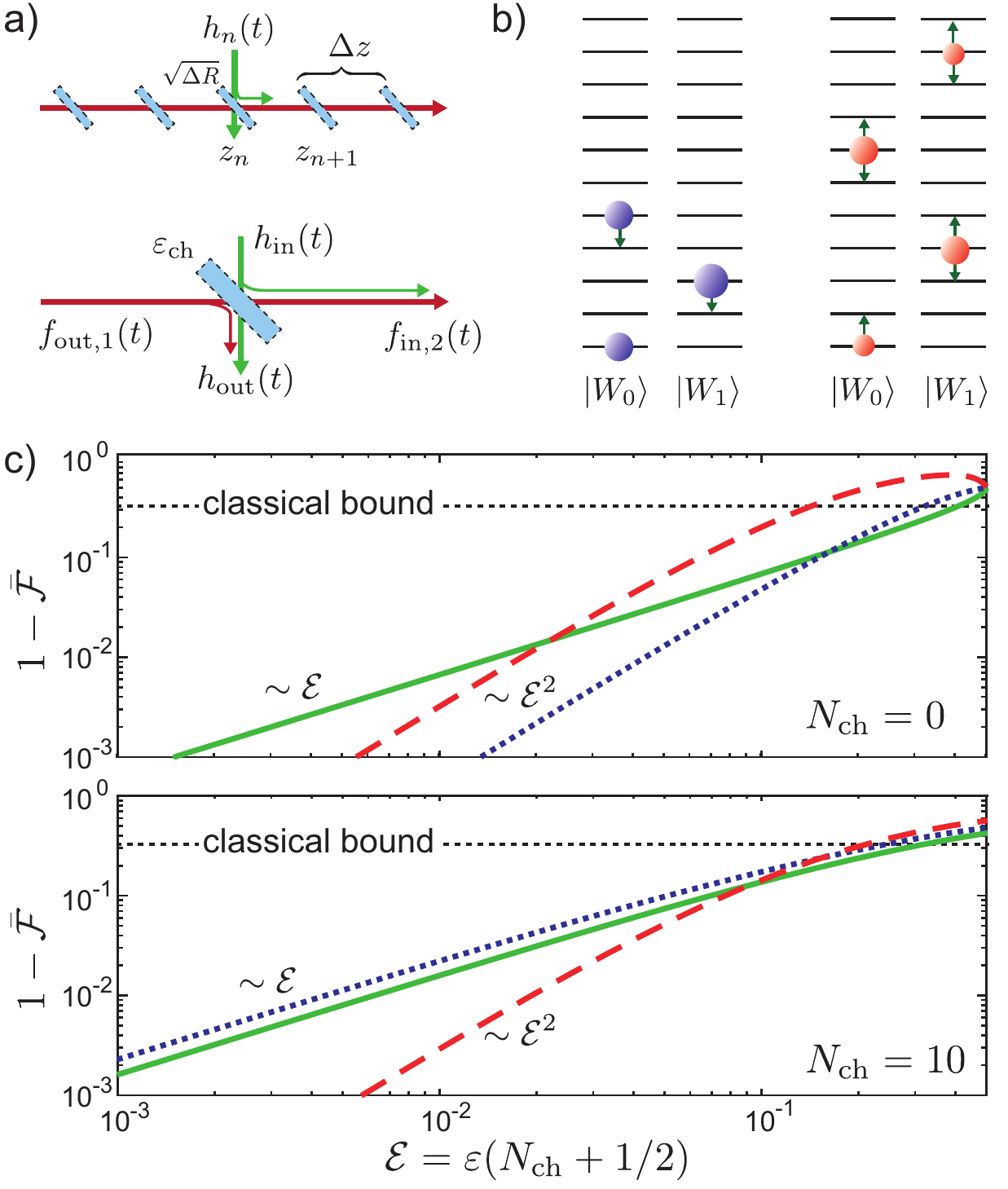}
      \caption{(a) Channel losses are modeled by a series of beam splitters with reflectance  $\Delta R=\Delta z/L_{\rm ab}$. For a homogeneous waveguide, the total loss can be taken into account by mixing the outgoing field of node 1 with a thermal field $h_{\rm in}$ on a beam splitter with reflectance $\varepsilon_{\rm ch}=1-e^{-L/L_{\rm ab}}$. (b) Illustration of the code words $|W_\sigma\rangle$ used for correcting single-photon loss errors (left) and for correcting single-photon loss and single-photon absorption errors (right). (c) The average transfer fidelity is shown for a state transfer protocol without error correction (green solid line), for the case when single-photon losses are corrected (blue dotted line), and for the case when both single-photon loss and single-photon absorption errors are corrected (red dashed line).}
      \label{Fig4}
\end{figure}

To include  the effect of propagation losses in a channel with  absorption length $L_{\rm ab}$, we divide the whole waveguide into small segments of length $
\Delta z$. The losses within one segment can be modeled by a beam splitter with reflectance $\Delta R=\Delta z/L_{\rm ab}$ [see Fig.~\ref{Fig4}(a)], which implies the following relation between the fields at $z_{n}$ and $z_{n+1}=z_n+\Delta z$, 
\begin{equation}
f_{\rm ch}\left(z_{n+1}, t+\frac{\Delta z}{v_{\rm g}}\right) =\sqrt{1-\Delta R}f_{\rm ch}(z_n, t)+\sqrt{\Delta R}h_{n}(t).
\end{equation}
Here $h_{n}(t)$ represents the thermal field entering through the second port of the beam splitter, which is characterized by the local thermal occupation number $N(z)$, i.e., $\langle h_{n}^\dag(t)h_{m}(t')\rangle=\delta_{n,m}\delta(t-t')N(z_n)$. By taking the continuum limit $\Delta z\rightarrow 0$, we then obtain the propagation equation~\cite{Habraken2012} 
\begin{equation}\label{eq:NoisePropagation}
\left(\frac{\partial}{\partial z}+ \frac{1}{v_{\rm g}}\frac{\partial}{\partial t}\right) f_{\rm ch}(z,t)= -\frac{f_{\rm ch}(z,t)}{2L_{\rm ab}}  + \frac{h(z,t)}{\sqrt{L_{\rm ab}}} ,
\end{equation}
where $h(z,t)$ is a continuous bosonic field, with variance $\langle h^\dag(z,t) h(z',t')\rangle = N(z) \delta(t-t')\delta(z-z')$. This equation results in the following modified propagation relation:
\begin{equation}
f_{\rm in,2}(t) =e^{-\frac{L}{2L_{\rm ab}}} f_{\rm out,1}(t)+ \frac{1}{\sqrt{L_{\rm ab}}}  \int_0^L dz \, e^{-\frac{(L-z)}{2L_{\rm ab}}}  h\left(z,t\right),
\end{equation}
where we have already absorbed all retardation times by an appropriate  redefinition of the field operators. For the following analysis, we focus on the case where the channel temperature is approximately uniform, $N(z)\simeq N_{\rm ch}$, and we can further simply this relation to
\begin{equation}\label{eq:Losses}
f_{\rm in,2}(t) =\sqrt{1-\varepsilon_{\rm ch}}[f_{\rm in}(t)+\sqrt{\gamma_1(t)}a_1(t)]+\sqrt{\varepsilon_{\rm ch}}h_{\rm in}(t).
\end{equation}
Here, $\varepsilon_{\rm ch}= 1-e^{-L/L_{\rm ab}}$ can be identified with  the total loss in the channel, and $h_{\rm in}(t)$ satisfies $\langle h_{\rm in}^\dag(t)h_{\rm in}(t')\rangle=\delta(t-t')N_{\rm ch}$.

By taking  into account  the influence of loss via relation~\eqref{eq:Losses} and assuming pulses long enough such that $A_2(t_f,t_0)\rightarrow 0$, the solution for $a_2(t=t_f)$ is 
\begin{equation}
\begin{split}
a_2(t_f) = & \sqrt{1-\varepsilon_{\rm ch}} \left[ T(t_f,t_0) a_1(t_0)+ F_2(t_f,t_0) \right] \\
-& \sqrt{\varepsilon_{\rm ch}} \int_{t_0}^{t_f}ds\, A_2(t_f,s)\sqrt{\gamma_2(s)}h_{\rm in}(s).
\end{split}
\end{equation}
This result already shows that, although in the presence of finite losses a perfect transfer is no longer possible, the transfer pulses obeying condition~\eqref{eq:DarkState} maximize $|T(t,t_0)|$, and therefore these pulses are still optimal. By now including small deviations from this condition, i.e., $T(t_f,t_0)=-\sqrt{1-\varepsilon_{\rm p}}$, the final result can be written in a compact form as (see Appendix~\ref{app:Solution})
\begin{equation}\label{eq:a2error}
\begin{split}
a_2(t_f) = & \sqrt{(1-\varepsilon_{\rm ch})(1-\varepsilon_{\rm p})}a_1(t_0)\\
& + \sqrt{\varepsilon_{\rm p}(1-\varepsilon_{\rm ch})} \,\bar f + \sqrt{\varepsilon_{\rm ch}}\, \bar h,
\end{split}
\end{equation}
where the effect of the integrated noise is now simply represented by two bosonic operators,  $\bar f$ and $\bar h$, obeying $[\bar f,\bar f^\dag]=[\bar h,\bar h^\dag]=1$ and $\langle \bar f^\dag \bar f\rangle =\langle \bar h^\dag \bar h\rangle=N_{\rm ch}$. 
From this exact relation between the Heisenberg operators of the nodes and the channel fields, we can reconstruct the reduced density matrix $\rho_{2}(t_f)$ after the transfer from the initial state of node 1, as detailed in Appendix~\ref{app:Reconstruction}. After expanding the result to lowest order in $\varepsilon\simeq \varepsilon_{\rm p}+\varepsilon_{\rm ch}$, we obtain
\begin{equation}\label{eq:KrausMap}
\rho_2(t_f) = \sum_{\ell=0,\pm} E_\ell \rho(t_0)E_\ell^\dag  +  \mathcal{O}(\varepsilon^2),
\end{equation}
where $E_+=\sqrt{\varepsilon N_{\rm ch}} a_2^\dag$ and $E_-=\sqrt{\varepsilon(N_{\rm ch}+1)} a_2$ are Kraus operators associated with the addition and subtraction of a  single photon and $E_0=1-\varepsilon[ (N_{\rm ch}+1/2) a_2^\dag a_2 +N_{\rm ch}/2]$ describes the distortion of the final state under the condition that no error has occurred. 

In summary, Eq.~\eqref{eq:KrausMap} shows  that, in a thermal network, various types of imperfections introduce both photon loss and photon absorption errors with a characteristic rate $\mathcal{E}=\varepsilon (N_{\rm ch}+1/2)$. In view of these thermally enhanced error rates, an important question is if and under which conditions such errors can be further reduced by supplementing quantum state transfer protocols with error-correction schemes. Here, again, the large Hilbert space of the intermediary oscillator is to our advantage. In general, a set of errors specified by the  Kraus map~\eqref{eq:KrausMap}, can be corrected by encoding the initial state in two code words $|\phi_0\rangle=\vert W_0\rangle$ and $|\phi_1\rangle=\vert W_1\rangle$, satisfying $\langle W_{\sigma} |E_{\ell}^{\dag} E_k| W_{\sigma'}\rangle=\alpha_{\ell k}\delta_{\sigma\sigma'}$~\cite{Bennett1996,Knill1997}. In addition, the measurement of an observable $O_{\rm M}$ must be able to distinguish the error words $|E_\sigma^\ell \rangle= E_\ell |W_\sigma\rangle/\sqrt{\alpha_{\ell \ell}}$ for different errors $\ell$, but without revealing information about $\sigma$. Then, for a measurement outcome $\ell$,  an appropriate unitary operation $R_\ell|E_\sigma^\ell \rangle= |W_\sigma\rangle$ can be implemented to recover the original states without affecting the original superposition.
 
To illustrate more specifically the application of error-correction schemes for the thermal state transfer problem at hand, we now consider two specific codes that have recently been introduced in Ref.~\cite{Michael2016} [see Fig.~\ref{Fig4}(b)]. The first code, $\vert W_0\rangle=(\vert0\rangle +\vert 4\rangle)/\sqrt{2}$, and $\vert W_1\rangle = |2\rangle$, corrects for photon loss only. This means that these two states remain orthogonal under the action of $E_-$ and the occurrence of an error can be detected by measuring $O_{\rm M}\equiv (a^\dag a)_{\rm mod 2}$, i.e., the photon number modulo 2. The second code uses the states $\vert W_0\rangle=(\vert0\rangle +\sqrt{3}\vert 6\rangle)/2$ and $\vert W_1\rangle = (\sqrt{3}\vert 3\rangle+\vert9\rangle)/2$. It corrects for both photon loss and photon absorption errors, which can be distinguished from each other by a measurement of the operator $O_{\rm M}\equiv (a^\dag a)_{\rm mod 3}$. We remark that such measurements represent generalized photon-number parity measurements as implemented in many pioneering cavity QED experiments 
with Rydberg atoms~\cite{Sayrin2011} or superconducting qubits~\cite{Sun2014,Heeres2015}.

In Fig.~\ref{Fig4}(c) we compare the performance of the two codes with respect to the noncorrected transfer of states $|0\rangle$ and $|1\rangle$. Note that, for this plot, it is assumed that all the measurements and recovery operations are implemented with perfect fidelity. As expected, for $N_{\rm ch}=0$, the first code corrects all single-photon losses, which leads to an improved scaling of the infidelity, $1-\bar{\mathcal{F}}$, from $\sim \mathcal{E}$ to $\sim \mathcal{E}^2$. Although the second code corrects photon losses as well, it performs much worse. This poor performance is due to the fact that the average excitation number $\langle W_\sigma | a^\dag a |W_{\sigma}\rangle=9/2$ is much higher than in the first encoding scheme where $\langle W_\sigma | a^\dag a |W_{\sigma}\rangle=2$ or in the case of the nonencoded states. This  enhances the effective rate for multiphoton errors which  are not corrected by these codes. At higher thermal occupation numbers, this picture changes. The correction of only photon losses is no longer enough, and we see that the first code performs even worse than a transfer without error correction. However, a more sophisticated encoding, even in very fragile states, as is the case for the second code, can still drastically improve state transfer fidelities once a threshold of $\mathcal{E}\lesssim 0.1$ is reached. This result demonstrates that even in the presence of weak losses, a fault-tolerant transfer of quantum states through thermal communication channels is possible.

\section{Applications for microwave networks}

 The ability to deterministically transfer quantum states through noisy communication channels can be crucial for a wide range of networking applications. Most importantly, it provides access to microwave-based quantum communication schemes already at temperatures of several kelvin, thereby drastically reducing the technical overhead that would otherwise be required to cool large networks down to millikelvin temperatures. In the field of superconducting quantum circuits, all the individual tools for implementing the proposed transfer and error-correction protocols have already been demonstrated and are the focus of ongoing experimental investigations. This includes the implementation of tunable decay rates~\cite{Yin13,Srinivasan14,Wenner2014,Pechal14,Andrews15,Flurin2015,Wulschner16,Pfaff2016} for releasing and catching the photons, coherent circulators  for unidirectional coupling \cite{Sliwa15,Kerckhoff15}, and techniques for preparing arbitrary superpositions of photon-number states~\cite{Heeres2015,Krastanov15} and number state readout~\cite{Ofek16}.

To estimate the distances over which a quantum state transfer through a thermal microwave network can potentially be achieved, we first consider a microwave channel  with an absorption loss of ~0.01 dB/m. This value has recently been measured~\cite{Kurpiers2016} for commercially available waveguides at a frequency of $\omega_0/(2\pi)\approx5$ GHz and $T=4$ K. It corresponds to an absorption length of roughly $L_{\rm ab}\approx 500$ m, which, together with a thermal occupation number of $N_{\rm ch}\approx 15$, translates into errors in the range of $\mathcal{E}\approx 0.03 - 0.15$ for transfer distances of $L=1-5$ m. This is already sufficient to deterministically transfer quantum states between two dilution refrigerators. However, for 3D superconducting microwave resonators, quality factors as high as $Q=10^6-10^8$~\cite{Reagor13} have been observed, which corresponds to an integrated propagation length of $L_{\rm ab}\approx cQ/\omega_0\approx 10^4\sim10^6$ m. Therefore, the intrinsic losses of optimized microwave waveguides can be substantially smaller, in which case transfer errors as low as  $\mathcal{E}\approx 0.0015 - 0.15$ over a distance of $L=100$ m become possible. These estimates demonstrate the huge potential of thermal microwave networks, ranging from fridge-to-fridge to intra-city quantum communication.

\section{Other experimental considerations}

In this section, we briefly comment on several other experimental issues, which are less fundamental but of relevance for the operation of larger networks, in particular, at nonzero temperature. This concerns, first of all, the incomplete decoupling of the local node from the thermal channel, meaning that $\gamma_i(t)\geq \gamma_{\rm min}$ cannot be switched off completely. As a consequence, the local oscillator will be in a thermal state with temperature $T\approx T_{\rm ch}$  at the beginning of the protocol, and  the thermal noise entering through the channel will degrade the encoding operation. The first issue can be solved by implementing another tunable decay channel into the zero-temperature bath with maximal rate $\gamma_{\rm c}\gg \gamma_{\rm min}$. This can be used to precool the local oscillator to a residual occupation number $N_{\rm res}=(\gamma_{\rm min}/\gamma_{\rm c})N_{\rm ch}\ll 1$. The error introduced during the encoding operation depends on the detailed implementation, but as long as it is sufficiently small, it can be approximately taken into account by a thermal Kraus map~\eqref{eq:KrausMap} acting on the ideal state $\rho(t_0)$. Both effects combined  lead to two additional  contributions in the error budget, i.e.,  $\varepsilon \rightarrow \varepsilon + \varepsilon_{\rm res} + \varepsilon_{\rm enc}$, where $\varepsilon_{\rm res}=\gamma_{\rm min}/\gamma_{\rm c}$ and $\varepsilon_{\rm enc}\approx \gamma_{\rm min}t_{\rm enc}$ for a time $t_{\rm enc}$ of the encoding operation $U_{\rm enc}$. By making the reasonable assumption that $t^{-1}_{\rm enc},\gamma_{\rm c}\gtrsim \gamma$, these additional errors are on the scale of the off/on ratio $\sim \gamma_{\rm min}/\gamma$. Note that flux-tunable~\cite{Wenner2014,Pierre2014} as well as  parametric~\cite{Flurin2015,Pfaff2016} couplers with maximal rates $\gamma/(2\pi)\sim 0.5-10$ MHz and off/on ratios of $~10^{-3}$ have already been experimentally demonstrated. 

\begin{figure}[t] 
  \centering
    \includegraphics[width=0.48\textwidth]{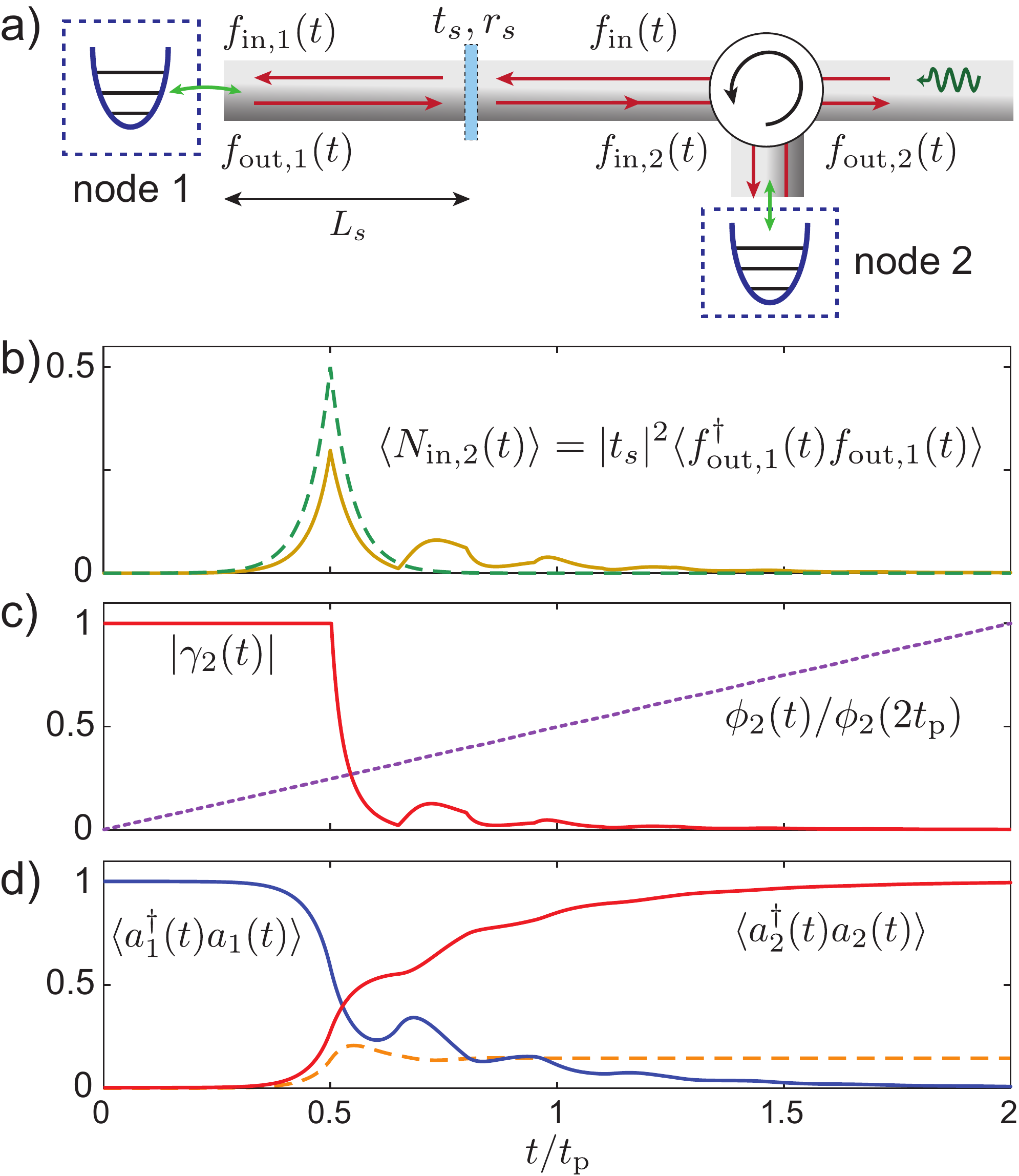}
      \caption{(a) The coherent backscattering of fields by defects or imperfect connections is modeled by a pointlike scatterer with reflectance $|r_s|^2=1-|t_s|^2$,  located at a distance $L_s$ from the first node. (b) Shape of the emitted wave packet (solid line) measured after the scatterer at the input of node 2. For this plot $\gamma_{1}(t)$, is the same as in Fig.~\ref{Fig3}(a), $\tau_s=2L_s/v_{\rm g}=0.15 t_{\rm p}$, $ t_{\rm p}=20\gamma^{-1}$, $r_s=\sqrt{1-t_s^2}=0.6$, and a detuning $\delta = \omega_1 - \omega_2 = \gamma$ between the two local oscillators has been assumed. The green dashed line shows the unperturbed wave packet for $r_s=0$. (c) Shape of the recovery pulse $\gamma_2(t)=|\gamma_2(t)| e^{-i2\phi_2(t)}$ with a final phase $\phi_2(2t_{\rm p})\simeq  2.03 \times  \delta t_{\rm p} $ obtained from a numerical optimization of the state transfer problem. (d) Plot of the resulting evolution of the population of each oscillator for an initial state $|\psi_q\rangle=|1\rangle$. The dashed line shows the population in the second oscillator for the original pulse $\gamma_2(t)$ given in Fig.~\ref{Fig3}(a).}
      \label{Fig5}
\end{figure}

When integrating multiple dilution refrigerators into larger networks, impedance mismatches  between the local nodes and the microwave channel, the integration of circulators or simply the presence of defects in the waveguide can lead to coherent backscattering, even without introducing loss. Such a situation is illustrated by the setting in Fig.~\ref{Fig5}(a), where a scattering element with reflectance  $R_s=|r_s|^2$ and transmittance $T_s=|t_s|^2$ is introduce between node 1 and node 2. This leads to a delayed backaction of the emitted field onto node 1, via the relation $f_{\rm in,1}(t)= t_s f_{\rm in}(t)+r_s f_{\rm out,1}(t-\tau_s)$, where $\tau_s=2L_s/v_{\rm g}$, and it significantly distorts the wave function at the receiver node, as shown in Fig.~\ref{Fig5}(b). However, in the absence of absorption, i.e., $|r_s|^2+|t_s|^2=1$, the requirements for implementing a perfect state transfer do not rely on a specific pulse shape but solely on $A_{1}(t_f,t_0)\rightarrow 0$ and on the dark state condition~\eqref{eq:DarkState} being fulfilled for all $t\in[t_0,t_f]$. Thus, by stepwise integrating the equations of motion for the network and adjusting   
\begin{equation}
  \sqrt{\gamma_2(t)} = -\sqrt{\gamma_1(t)}\frac{A_1(t,t_0)}{ T(t,t_0)},
\end{equation} 
at each point  in time, a perfect recovery pulse $\gamma_2(t)$ can also be found numerically for more complicated propagation relations, semidirectional and bidirectional channels, etc. Note that, in general, this construction requires that $\gamma_2(t)=|\gamma_2(t)|e^{-i2\phi_2(t)}$ must be tunable in phase and amplitude, as is the case for parametric couplers~\cite{Flurin2015,Pfaff2016}. A time-dependent $\phi_2(t)$ can further be used to compensate for small frequency shifts between the local resonators, which is another source of imperfections for couplers without this tunability~\cite{Sete2015}. An example for a  numerically optimized recovery pulse and the resulting transfer dynamics is shown in Figs.~\ref{Fig5}(c) and \ref{Fig5}(d).

Finally, we emphasize that throughout this work we have assumed that at the beginning of the transfer the channel field is in equilibrium at temperature $T_{\rm ch}$. This is a reasonable assumption for cases where the two nodes of interest are part of a much larger connected network in contact with a thermal reservoir. However, given the rather long rethermalization length and time scales, $L_{\rm ab}$ and $L_{\rm ab}/v_{\rm g}$ [see Eq.~\eqref{eq:NoisePropagation}], the effective occupation number $N_{\rm in}=\langle \bar f^\dag \bar f\rangle$ characterizing the incoming field $f_{\rm in}(t)$ can be both much smaller than $N_{\rm ch}$ [e.g., $ N_{\rm in}\sim (L/L_{\rm ab})N_{\rm ch}$, if the input channel is connected instead to a cryogenic reservoir] and also much higher [e.g., if parts of the network are affected by nonequilibrium (technical) noise]. In contrast, the noise associated with absorption,
\begin{equation}
\langle \bar h^\dag \bar h\rangle = \frac{1}{\varepsilon_{\rm ch} L_{\rm ab}}\int_0^L dz \, e^{-\frac{(L-z)}{L_{
\rm ab}}}  N_{\rm mat}(z),
\end{equation}
is mainly determined by the temperature $T_{\rm mat}(z)$ of the waveguide material along the channel, i.e., $N_{\rm mat}(z)=(e^{\hbar\omega_0/k_BT_{\rm mat}(z)}-1)^{-1}$. These considerations show that apart from the routing of quantum information, controlling the flow of noise fields may by itself become a relevant optimization problem in larger microwave networks. Independent of the precise operation strategy, coherent noise cancellation as discussed in this work can be very beneficial since it avoids many additional switching and cooling elements and can tolerate noise levels far beyond what is acceptable for qubit networks.

\section{Conclusions and outlook}

In summary, we have shown how quantum states can be transmitted with close to unit fidelity through a thermal channel. The key ingredient, namely, the use of an intermediary oscillator as a controllable port between the local qubit and the quantum communication channel, not only provides the necessary degree of linearity for a coherent cancellation of noise, it also enables the implementation of protocols for correcting the residual photon loss and absorption errors under realistic conditions. This combination opens the new possibility of building robust microwave quantum networks on medium and large scales. While the coherent cancellation of noise would already be very beneficial for continuous variable quantum communication~\cite{DiCandia2015} and key distribution~\cite{Grosshans2003,Weedbrook2010} schemes in this frequency range, the protocol enables, in particular, a direct deterministic exchange of quantum information over thermal microwave channels at rates of about $10^6-10^8$ qubits per second. Such capabilities are very challenging to implement in the optical domain and may have a wide-ranging impact on future quantum communication strategies.

\vspace{0.5cm}
\emph{Note added.} A very similar protocol has been described in a parallel work by B. Vermersch {\it et al.}~\cite{Vermersch2016}.

\acknowledgements
 We thank F. Deppe, S. Habraken, and J. Majer for valuable input. 
This work was supported by the Austrian Science Fund (FWF) through SFB FOQUS F40 and the START Grant No. Y 591-N16 and by the European Commission through the Marie Sklodowska-Curie Grant No. IF 657788.
L.J. acknowledges support from the ARL-CDQI, ARO (Grant No. W911NF-14-1-0011, W911NF-14-1-0563), ARO MURI (Grant No. W911NF-16-1-0349), NSF (Grant No. EFMA-1640959), AFOSR MURI (Grant No. FA9550-14-1-0052 and No. FA9550-15-1-0015), Alfred P. Sloan Foundation (Grant No. BR2013-049), and Packard Foundation (Grant No. 2013-39273).


\appendix
\section{General solutions for the Heisenberg operators}\label{app:Solution}
In this appendix, we summarize the general solutions for the Heisenberg operators $a_i(t)$. We start with the quantum Langevin equation for $a_1(t)$ given in Eq.~\eqref{eq:QLE}, which can be directly integrated to obtain 
\begin{equation}
a_1(t)= A_1(t,t_0)a_1(t_0) - \int_{t_0}^t ds   \, \sqrt{\gamma_1(s)} A_1(t,s) f_{\rm in}(s), 
\end{equation}
where $\frac{d}{dt} A_1(t,t_0)=-\frac{\gamma_1(t)}{2}  A_1(t,t_0)$ and $A_1(t_0,t_0)=1$. By using this result and including channel loss via relation~\eqref{eq:Losses}, the integration of the resulting quantum Langevin equation for $a_2(t)$ gives
\begin{equation}
\begin{split}
a_2(t)= &A_2(t,t_0)a_2(t_0) + \sqrt{1-\varepsilon_{\rm ch}} T(t,t_0) a_1(0) \\
&+  \sqrt{1-\varepsilon_{\rm ch}}F_2(t,t_0)\\
&- \sqrt{\varepsilon_{\rm ch}} \int_{t_0}^{t}ds\, A_2(t,s)\sqrt{\gamma_2(s)}h_{\rm in}(s), 
\end{split}
\end{equation} 
where $T(t,t_0)$ is defined in Eq.~\eqref{eq:T} and 
\begin{equation}\label{eq:App:Noise}
\begin{split}
F_2(t,t_0)&=-\int_{t_0}^t dt' A_2(t,t')\sqrt{\gamma_2(t')} f_{\rm in}(t')\\
& +\int_{t_0}^t dt' A_2(t,t')\sqrt{\gamma_1(t')\gamma_2(t')} A_1(t',t_0)\\
&\times \int_{t_0}^{t'} dt'' A_1(t_0,t'') \sqrt{\gamma_1(t'')}f_{\rm in}(t'').
\end{split}
\end{equation}
We can write this integral as $ F_2(t,t_0)= \int_{t_0}^t [ X(t')- \dot Y(t') Z(t')]dt'$, where $X(t')$, $\dot Y(t')$, and $Z(t')$ represent the individual terms in each line of Eq.~\eqref{eq:App:Noise}. Then, after integrating by parts, i.e., $F_2(t,t_0)=\int_{t_0}^t  X(t')dt'  -  Y(t)Z(t)+\int_{t_0}^t dt'  Y(t') \dot Z(t')$, we obtain the simpler expression $ F_2(t_f,t_0)= \int_{t_0}^t ds\, D_2(t,s) f_{\rm in}(s)$, with $D_2(t,s)$ given in Eq.~\eqref{eq:D2}.

From the commutation relation $[a_2(t),a_2^\dag(t)]=1$, we obtain the general result
\begin{equation}
\begin{split}
1&= |A_2(t,t_0)|^2 + (1-\varepsilon_{\rm ch}) T^2(t,t_0) \\
&+  (1-\varepsilon_{\rm ch}) \int_{t_0}^t ds \,D^2_2(t,s) + \varepsilon_{\rm ch} [1- |A_2(t,t_0)|^2]. 
\end{split}
\end{equation}
By assuming imperfect pulses with $T^2(t_f,t_0)=1-\varepsilon_{\rm p}$, but long enough such that $A_2^2(t_f,t_0)\rightarrow0$, this relation allows us to identify the correct prefactors for the noise operators $\bar f$ and $\bar h$ in Eq.~\eqref{eq:a2error}, independent of the exact pulse shape.

\section{Reconstruction of the density operator}\label{app:Reconstruction}
In this appendix we derive the explicit result for the final state of the second node $\rho_{2}(t_{f})=\sum_{r,r'=0}^{\infty}[\rho_{2}(t_{f})]_{r,r'}|r\rangle\langle r'|$ for a given initial state of the first node $\rho_{1}(t_{0})=\sum_{n,n'=0}^{\infty}[\rho_{1}(t_{0})]_{n,n'}|n\rangle\langle n'|$~\cite{WangNJP12,Sete2015}.
According to Eq.~\eqref{eq:a2error}, we can write
\begin{equation}
a_{2}\left(t_{f}\right)=\sqrt{1-\varepsilon}a_{1}\left(t_{0}\right)+\sqrt{\varepsilon}b,\label{eq:mapping}
\end{equation}
where $\varepsilon=\varepsilon_{{\rm ch}}+\varepsilon_{{\rm p}}-\varepsilon_{{\rm ch}}\varepsilon_{{\rm p}}$
and we have combined the two noise fields into a single bosonic operator 
\begin{equation}
b=\frac{1}{\sqrt{\varepsilon}}\left(\sqrt{\varepsilon_{{\rm p}}(1-\varepsilon_{{\rm ch}})}\,\bar{f}+\sqrt{\varepsilon_{{\rm ch}}}\,\bar{h}\right),
\end{equation}
obeying $[b,b^{\dag}]=1$ and $\langle b^\dag b\rangle=N_{\rm ch}$. Note that this assumes equal temperatures for the original fields, $f_{\rm in}$ and $h_{\rm in}$.  
Given that the mode $b$ is a mixture of different number states $|k\rangle$
with probability $p_{k}$, the mapping of Eq.~\eqref{eq:mapping} corresponds
to a completely positive and trace preserving map from the first node to the second node with Kraus operator
representation
\begin{equation}
\rho_{2}\left(t_{f}\right)=\sum_{k=0}^{\infty}\sum_{q=0}^{\infty}K_{k,q}\rho_{1}\left(t_{0}\right)K^\dag_{k,q}.
\end{equation}
The Kraus operators are 
\begin{equation}
K_{k,q} =\sum_{r}\sqrt{p_{k}}\mathcal{K}\left(r,r+q-k,k\right) |r\rangle_{2}\langle r+q-k|_{1},
\end{equation}
with
\begin{equation}
\begin{split}
\mathcal{K}(r,n,k)=\sum_{i=0}^{n} &(-1)^{n-i}\varepsilon^{\frac{n+r-2i}{2}}(1-\varepsilon)^{\frac{k-r+2i}{2}}\\
&\times \binom{n}{i}\binom{k}{r-i}\sqrt{\frac{\binom{n+k}{n}}{\binom{n+k}{r}}}\label{eq:K-function}
\end{split}
\end{equation}
for $r\le n+k$ and $\mathcal{K}(r,n,k)=0$ otherwise. 
The binomial
coefficients are $\binom{n}{m}=\frac{n!}{m!\left(n-m\right)!}$ for
$0\le m\le n$ and $\binom{n}{m}=0$ otherwise. 
Finally, we obtain the expression in terms of the density matrix elements in the number basis,
\begin{equation}
[\rho_{2}(t_{f})]_{r,r'}=\sum_{k=0}^{\infty}p_{k}\sum_{n,n'=0}^{\infty}G(r,r',n,n';k)[\rho_{1}(t_{0})]_{n,n'},
\end{equation}
with 
\begin{equation}
G(r,r',n,n';k)=\mathcal{K}(r,n,k)\mathcal{K}(r',n',k),
\end{equation}
and $\mathcal{K}(r,n,k)$ as defined in Eq.~\eqref{eq:K-function}.


\bibliographystyle{apsrev4-1-PRX}  
\bibliography{ThermTrans_references}

\end{document}